\documentclass[11pt]{article}
\usepackage{geometry}                
\usepackage{graphicx}
\usepackage{amsmath}
\usepackage{amssymb}

\newcommand{\ket}[1]{\left|#1\right\rangle}

\usepackage{fancyheadings}

\numberwithin{equation}{section}
\title{State Space for Planar Majorana Zero Modes}
\author{R. Jackiw$^1$ and  S.-Y. Pi$^2$\\[1.5ex]
\small\itshape $^1$ Physics Department, MIT, Cambridge MA 02139\\[.5ex]
\small\itshape $^2$ Physics Department, Boston University, Boston MA 02215}
\date{}                                           

\begin{document}
\maketitle
\thispagestyle{fancy}
\begin{abstract}
Zero modes arising from a planar Majorana equation in the presence of $N$ vortices require an $\mathcal{N}$-dimensional state-space, where $\mathcal{N} = 2^{N/2}$ for $N$ even and $\mathcal{N} = 2^{(N + 1)/2}$ for $N$ odd. The mode operators form a restricted $\mathcal{N}$-dimensional Clifford algebra.
\end{abstract}
\section{Introduction}
Majorana fermions are of central  interest for recent research on particle physics, cosmology, condensed matter physics.

In Nature there are examples of charge neutral bosons that are their own anti-particles, while there are no such known examples of charge neutral fermions. Majorana modified the Dirac equation, speculating that the neutrino might be its own anti-particle and could be described by the Majorana equation. Discovery of neutrino oscillations makes Majorana's idea compelling, since massive neutrinos, with individual lepton number non-conservation, are naturally described by Majorana's equation. 

Majorana fermions also occur as hypothetical particles in supersymmetric theories and  in models for dark matter.

Recent development in condensed matter physics predicts occurrence of Majorana fermions in exotic superconductors. We consider a system in which a topological insulator is in contact with a s-wave superconductor. Due to the proximity effect, Cooper pairs tunnel across the surface and interface excitations are described by the Majorana equation (except that the geometry is planar and a chemical potential is present) \cite{FuPhys08}.

When a Majorana equation for a chargeless fermion field governs such a topological superconductor, there may arise zero-energy eigenmodes on a topologically non-trivial background configuration. In a planar geometry, an $N$-vortex background leads to $N$ zero modes \cite{jack1981}. Due to the fact that these Majorana zero modes are topologically protected and satisfy non-Abelian statistics, they play an important role in recent research on topological quantum computing \cite{Nayak:2008zza}.

The question arises what is the dimensionality and detailed structure of the state space that  accommodates these zero modes. Analysis of the system leads to a result for even $N$ \cite{Nayak:2008zza}. The number of states $\mathcal{N}$  is 
\begin{equation}
\mathcal{N} = 2^{N/2} \quad (N \ \text{even}) .
\label{jackiw1.1}
\end{equation}
The situation for odd $N$ is mostly ignored, because in the condensed matter community attention is focussed on pairs of Majorana modes, with each pair acting as a conventional Dirac fermion. Nevertheless, it should be possible to understand the structure for arbitrary $N$, both even and odd, and to give a mathematically unified description. This is what we provide here, thereby extending our previous $N=1$ analysis, which we now summarize \cite{Chamon:2010ks}.

\section{\underline{$N=1$}}
A Majorana fermion field $\Psi$, moving on a plane in the presence of a single vortex, possesses the mode expansion
\begin{equation}
\Psi = \sum_{E>0} \ (a_E\, e^{-iEt}\, u_E + a^\dagger_E\, e^{iEt}\, C\, u^\ast_E) + a\, u_{E=0}
\label{jackiw2.1}
\end{equation}
The $u_E$ are the $E\ne0$ mode functions governed by the creation/annihilation operators $a^\dagger_E/a_E$, while $u_{E=0}$ and $a= a^\dagger$  serve the same purpose for the zero mode. $C$ is the conjugation matrix and $\Psi$ satisfies the Majorana condition $C\Psi^\dagger = \Psi$.

The anti-commutation relations obeyed by $\Psi$ require that the zero mode operator $a$ satisfies
\begin{subequations}
\begin{equation}
\{a, a^\dagger\} = 1.
\label{jackiw2.2a}
\end{equation}
But $a$ is also Hermitian, therefore,
\begin{equation}
a^2 = 1/2\, .
\label{jackiw2.2b}
\end{equation}
\end{subequations}
Apart from a factor $\frac{1}{\sqrt{2}}, a$ is unitary, and its action is effectively norm preserving. The state space of $a$ can take two different forms (i) and (ii), as is explained in \cite{Chamon:2010ks}.
\begin{itemize}
\item[(i)]
It may be that $a$ is diagonal on two states $\ket{0 \pm}$.
\begin{equation}
a \ket{0 \pm} = \pm \frac{1}{\sqrt{2}} \ket{0 \pm}
\label{jackiw2.3}
\end{equation}
There are two ground states. Two towers of excited states are constructed with repeated applications of $a^\dagger_E$. No operator connects the two. Fermion parity is broken because $a$ is a fermionic operator; therefore, the two sides of \eqref{jackiw2.3} have opposite fermion parity. Like in spontaneous symmetry breaking, a vacuum $\ket{0 +}$ or $\ket{0 -}$ must be chosen and no tunneling connects to the other ground state. Evidently we have two 1-dimensional realizations.

\item[(ii)] Alternatively we can have one 2-dimensional realization: the vacuum is doubly degenerate $\ket{1}, \ket{2}$ and  $a$ connects the two vacua, one bosonic the other fermionic.
\begin{eqnarray}
a \ket{1} &=& \frac{1}{\sqrt{2}}\ \ket{2}\nonumber\\
a\ket{2}&=& \frac{1}{\sqrt{2}}\  \ket{1}
\label{jackiw2.4}
\end{eqnarray}
Fermion parity is preserved, and mixing between $\ket{1}$ and $\ket{2}$ is not allowed. Repeated action of $a^\dagger_E$ creates two towers of states, connected by $a$.
\end{itemize}

Ref  \cite{Chamon:2010ks} provides an argument in favor of the second, fermion parity preserving realization. The argument begins by considering a background of a widely separated vortex/anti-vortex pair. There are no zero-modes; rather there are two modes $u_{\pm \varepsilon}$, one with (small) positive energy $\varepsilon$, another with negative energy $-\varepsilon$. There is no ambiguity: the positive energy mode enters with an annihilation operator, $a_\varepsilon\, u_\varepsilon$ the negative energy mode with a creation operator, $a^\dagger_\varepsilon\, u_{-\varepsilon}$. The two low lying states are the vacuum $\ket{\Omega}, a_\varepsilon \, \ket{\Omega}=0$ and the filled state $\ket{\varepsilon} = a^\dagger_\varepsilon \, \ket{\Omega}$. As the distance between the vortex/anti-vortex ranges to infinity, $\varepsilon \to 0$ and the mode functions  $u_{\pm \varepsilon}$ collapse into one, leaving
\begin{equation}
\begin{array}{c|c}
\left(a_\varepsilon + a^\dagger_\varepsilon\right) \!\! & 
    \!\! u_0 = a u_0\, .\\[1ex]
     &\!\! \! \!\raisebox{-1ex}{$\varepsilon_{= 0}$} \hspace{2.25em}
      \end{array}
\label{jackiw2.5}
\end{equation}
The two low lying states survive as our previously defined $\ket{1} \text{ and } \ket{2}: \ket{\Omega}\ \raisebox{-1.25ex}{{$\small\overrightarrow{\varepsilon \to  _{0}}$}}\ \ket{1}, \ket{\varepsilon}\ \raisebox{-1.25ex}{{$\small\overrightarrow{\varepsilon \to  _{0}}$}}\ \ket{2}$, with $a$ connecting them.

This reproduces the two-dimensional, fermion parity preserving realization. An explicit representation is provided by a Pauli matrix.
\begin{equation}
a = \frac{\sigma_1}{\sqrt{2}}, \ \ket{1} = \binom{1}{0}, \ket{2} = \binom{0}{1}
\label{jackiw2.6}
\end{equation}
We  adopt this representation and extend it to higher $N$. 

(An explicit realization of the two, 1-dimensional, fermion parity violating representation is obtained by setting $a = \frac{\sigma_3}{\sqrt{2}}, \ \ket{0+} = \binom{1}{0}, \ket{0-} = \binom{0}{1}$.)

\section{\underline{$N=2$}}
With two vortices, we have two zero modes with Hermitian mode operators $a$ and $b$, satisfying 
\begin{equation}
a^2 = 1/2,\, b^2 = 1/2,\, ab +ba =0 \, .
\label{jackiw3.1}
\end{equation}
A two-dimensional realization is again possible.
\begin{subequations}\label{jackiw3.2}
\begin{eqnarray}
a\ket{1} &=& \frac{\alpha}{\sqrt{2}} \ \ket{2}\nonumber\\[1ex]
a\ket{2} &=& \frac{1}{\alpha\sqrt{2}}\  \ket{1}\label{jackiw3.2a}\\[4ex]
b\ket{1} &=& \frac{\beta}{\sqrt{2}}\  \ket{2}\nonumber\\[1ex]
a\ket{2} &=& \frac{1}{\beta\sqrt{2}}\ \ket{1}
\label{jackiw3.2b}
\end{eqnarray}
\end{subequations}
with $|\alpha| = |\beta| = 1$. The vanishing anti-commutator of $\{a,b\}$ requires
\begin{equation}
\frac{\alpha}{\beta} + \frac{\beta}{\alpha} = 0\, .
\label{jackiw3.3}
\end{equation}
This is solved by $\alpha=1, \beta= i$ (apart from irrelevant phases). Thus the $N=2$ case regains the general, even-$N$ formula \eqref{jackiw1.1}: $\mathcal{N}=2 \ \text{ at }\ N = 2$. A concrete realization is given with $\ket{1}, \ket{2}$ as in \eqref{jackiw2.6} and 
\begin{equation}
a = \frac{\sigma_1}{\sqrt{2}},\, \ b = \frac{\sigma_2}{\sqrt{2}}\, .
\label{jackiw3.4}
\end{equation}
Now the explicit realization uses two Pauli matrices, $\sigma_1$ and $\sigma_2$, acting on the same space as in \eqref{jackiw2.6}.

\section{\underline{$N=3$}}
With three vortices, we first try a two dimensional representation for the three zero-mode operators: $a, b$ act as in \eqref{jackiw3.2} and the third $c$, is assumed to satisfy
\begin{eqnarray}
c \ket{1} &=& \frac{\gamma}{\sqrt{2}}\ \ket{2} ,\nonumber\\[1ex]
c \ket{2} &=& \frac{1}{\sqrt{2} \gamma}\ \ket{1} ,\label{jackiw4.1}\\[1ex]
|\gamma| &=& 1\, .\nonumber
\end{eqnarray}
The anti-commutators between $a,b,c$ lead to the conditions
\begin{eqnarray}
\frac{\alpha}{\beta} + \frac{\beta}{\alpha} = 0\nonumber\\[1ex]
\frac{\beta}{\gamma} + \frac{\gamma}{\beta} =0\nonumber\\[1ex]
\frac{\gamma}{\alpha} + \frac{\alpha}{\gamma} = 0
\label{jackiw4.2}
\end{eqnarray}
But these three conditions are inconsistent; a two dimensional fermion parity preserving realization is impossible for $N=3$. 

The same conclusion emerges if one attempts to realize $c$ as $\frac{\sigma_3}{\sqrt{2}}$. While this anti-commutes with $a$ and $b$, its action on the states $\ket{1}, \ket{2}$, is diagonal; they are eigenstates of $\sigma_3$. Consequently fermion parity is not preserved.

Evidently we must use a four-dimensional representation, which is obtained by iterating the two-dimensional one. The explicit realization makes use of Dirac matrices.
\begin{equation}
\boldsymbol{\alpha}  =\left(
						\begin{array}{cc}
						0  & i {\boldsymbol \sigma}  \\
						- i {\boldsymbol \sigma}  & 0 
						\end{array}
						\right), \ \ \beta =  \left(
						\begin{array}{cc}
						 0 & I \\
						 I & 0
						\end{array}
						\right)
						\label{jackiw4.3}
\end{equation}
We identify $(a, b,c)$ with any three matrices chosen from  $\frac{\boldsymbol \alpha}{\sqrt{2}}, \frac{\beta}{\sqrt{2}}$ and the four four-component states are
\begin{equation}
(1,0,0,0)^T, (0,1,0, 0)^T, (0,0,1,0)^T, (0,0,0,1)^T\, .
\label{jackiw4.4}
\end{equation}.
The anti-commutators of the mode operators are satisfied thanks to the algebra of the Dirac matrices, while their action on the states produces orthogonal states so that  fermion parity holds. (For this reason we do not use the Dirac matrix $\left(
						\begin{array}{cc}
						 I & 0 \\
						0 & -I
						\end{array}
						\right)$, which also anti-commutes with the others, but the four states are its eigenstates, which contradicts fermion parity preservation.)

\section{\underline{$N=4$}}
The four zero mode operators $(a, b, c, d)$ are now realized by $\frac{\boldsymbol \alpha}{\sqrt{2}}$ and $\frac{\beta}{\sqrt{2}}$, acting on the same four states as for $N=3$. This verifies the general formula \eqref{jackiw1.1}: $\mathcal{N} =4$ for $N=4$.

\section{\underline{Larger $N$}}
The pattern is now clear. For odd $N$ the even-$N$ formula \eqref{jackiw1.1} is changed to 
\begin{equation}
\mathcal{N} = 2^{\frac{N+1}{2}} \qquad (N \text{ odd}) ,
\label{jackiw6.1}
\end{equation}
while the mode operators form a Clifford algebra, realized by $\mathcal{N} \times \mathcal{N}$ Dirac matrices for both odd and even $N$. Adjacent odd and subsequent even cases have the same dimensionality. Passing beyond the adjacent odd/even pair into the next odd case requires doubling the Dirac matrices in an off-diagonal manner. (Diagonal Dirac matrices [in the Cartesian basis like \eqref{jackiw4.4}] do not arise in this context, because their action would not produce an orthogonal state, thus breaking fermion parity.) Eq. \eqref{jackiw6.1} is in agreement with $N=1, 3 \Rightarrow \mathcal{N} = 2, 4$ respectively.

\section{\underline{Conclusion}}
Beginning with a single vortex and zero mode, we adopted the two dimensional state space \eqref{jackiw2.4} on which fermion parity is conserved. This leads to  the familiar even-$N$ state counting formula \eqref{jackiw1.1}; and implies \eqref{jackiw6.1} for odd $N$. 

Formula \eqref{jackiw6.1} may be viewed as describing odd $(N)$ vortices plus one more ``at infinity," resulting in even number $(N+1)$ vortices, governed now by the even number formula \eqref{jackiw1.1}. We have used algebraic arguments to count states. With the inclusion of the phantom vortex at infinity it should be possible to construct a braiding argument for our result.

For the single vortex, we rejected the one-dimensional, fermion parity violating realization \eqref{jackiw2.3}. It is an open question whether this alternative realization has a role in physical theory. We note that within supersymmetry, fermion parity violation is a recognized and accepted phenomenon \cite{Losev:2001uc}.

We acknowledge helpful conversations with F. Haldane, I. Herbut, R. Lutchyn and Shankar. This research was supported by DOE contracts DE-FG02-05ER41360 (RJ) and DEF-91ER40676 (S-YP), and was performed at the NSF-supported Aspen Center for Physics.

\end{document}